\documentclass[12pt,letterpaper]{article}

\usepackage{amsmath}
\usepackage{amsfonts}
\usepackage{amssymb}
\usepackage{graphicx}
\usepackage{color}
\newtheorem{theorem}{Theorem}
\usepackage{graphicx}
\usepackage{epstopdf}
\usepackage{amsmath}
\usepackage{amsfonts}
\usepackage{amssymb}
\usepackage{color}
\usepackage{mathrsfs}
\usepackage{color}
\usepackage{enumerate}
\usepackage{cancel,dsfont} 
\pdfoutput=1
\usepackage{bbm}

\usepackage{mathrsfs}
\usepackage{amsfonts, amssymb}
\usepackage{graphicx}
\usepackage{psfrag}
\usepackage[usenames,dvipsnames,svgnames,table]{xcolor}
\usepackage{enumerate}

\usepackage{xcolor,url}
\usepackage{cancel,dsfont}

\usepackage[colorlinks,bookmarks]{hyperref}
\definecolor{linkblue}{rgb}{0,0,0.8}
\definecolor{linkgreen}{rgb}{0,0.5,0}

\hypersetup{pdfpagemode=UseNone, pdfstartview=FitH, linkcolor=linkblue,
            citecolor=linkgreen, urlcolor=linkblue}

\newcommand\eea{\end{eqnarray}}
\newcommand\bea{\begin{eqnarray}}

\def\be{\begin{equation}}
\def\ee{\end{equation}}

\def\nn{\nonumber}

\newtheorem{definition}{Definition}

\begin{document}

\begin{center}

{\Large \bf  Asymptotic Behavior of Cosmologies \\[0.4cm] with $\Lambda >0$ in 2+1 Dimensions}
\\[0.7cm]
{\large  Paolo Creminelli${}^{1}$, 
Leonardo Senatore${}^{2}$ , Andr\'as Vasy${}^{3}$ }
\\[0.7cm]

{\normalsize { \sl $^{1}$Abdus Salam International Centre for Theoretical Physics\\ Strada Costiera 11, 34151, Trieste, Italy\\[0.1cm]
IFPU - Institute for Fundamental Physics of the Universe,\\
Via Beirut 2, 34014, Trieste, Italy}}\\
\vspace{.3cm}

{\normalsize { \sl $^{2}$ Stanford Institute for Theoretical Physics,\\ Physics Department, Stanford University, Stanford, CA 94306}}\\
\vspace{.1cm}

{\normalsize { \sl 
Kavli Institute for Particle Astrophysics and Cosmology, \\
Physics Department and SLAC, Stanford University, Menlo Park, CA 94025}}\\
\vspace{.3cm}

{\normalsize { \sl $^{3}$ Mathematics Department,\\ Stanford University, Stanford, CA 94306}}\\
\vspace{.1cm}

\vspace{.3cm}

\end{center}

\vspace{.8cm}

\hrule \vspace{0.3cm}
{\small  \noindent \textbf{Abstract}} \\[0.3cm]
\noindent 
We study, using Mean Curvature Flow methods, 2+1 dimensional cosmologies with a positive cosmological constant and matter satisfying the dominant and the strong energy conditions. If the spatial slices are compact with non-positive Euler characteristic and are initially expanding everywhere, then we prove that the spatial slices reach infinite volume, asymptotically converge on average to de Sitter and they become, almost everywhere, physically indistinguishable from de Sitter. This holds true notwithstanding the presence of initial arbitrarily-large density fluctuations and the formation of black holes.
 
 \vspace{0.3cm}
\hrule


 \vspace{0.3cm}

\section{Introduction and Set-up}
The question of how likely it is for inflation to start has been discussed for many years, almost since its original formulation. In fact, while it is clear that inflation does inflate away inhomogeneities once it has started, its beginning seems to require approximate homogeneity over an inflationary Hubble patch, {\em i.e.}~a volume whose linear size is of the order of the Hubble radius of the homogeneous inflationary solution (this coincides with the de Sitter radius, if the inflationary phase is approximated with a de Sitter solution). This initial condition appears very unlikely, and gives rise to what is the so-called `initial patch problem' (see for example~\cite{Ijjas:2015hcc}). This well-motivated expectation has been recently shown to be incorrect for some interesting and non-trivial physical reasons. Ref.~\cite{East:2015ggf,Kleban:2016sqm} (see also~\cite{barrow1985closed}) have used a combination of numerical and analytical techniques to show that, for most of the three-dimensional topologies, and for a huge class of inhomogenous and anisotropic initial conditions, inflation will always start somewhere on the manifold, notwithstanding the formation of localized black-holes. Refs.~\cite{Clough:2016ymm,Clough:2017efm} have subsequently directly verified this on an extended range of initial conditions~\footnote{This indeed confirms what was expected in~\cite{East:2015ggf}: once the initial conditions are taken from a general enough class, since the system is very non-linear and even forms singularities, it explores the whole class of non-linear solutions after a Hubble time.}.
These results were made possible by recent developments in numerical relativity that allow to handle singularities and horizons~\cite{Pretorius:2005gq}, and, in Mathematics, with theorems such as the Thurston Geometrization Classification (see \cite{besse1987einstein} Theorem 4.35 and \cite{thurston1997three,10.2307/2152760}), and techniques like the Mean Curvature Flow (MCF)~(see, for example~\cite{gerhardtbook}).

At this point, it appears clear that the inflationary `initial patch problem', stating that it is unlikely for inflation to start somewhere out of most inhomogeneous initial conditions, is false. To argue that this is still a problem, requires one to find a reason why very peculiar initial conditions are preferred. 

In more detail, Ref.~\cite{Kleban:2016sqm} used the Thurston Geometrization Classification and MCF to show that, if there is a positive cosmological constant\footnote{In this paper the inflaton potential is replaced by a cosmological constant. Therefore we do not address issues like how likely it is for the inflaton to start sufficiently far away from the minimum of the potential.  In numerical studies \cite{East:2015ggf,Clough:2016ymm,Clough:2017efm} it was observed that once the initial conditions of the inflaton field are entirely contained within the inflationary part of the potential, it behaves effectively like a cosmological constant. This justifies and motivates our approximation. }, and matter satisfies the weak energy condition, all initially-expanding 3+1 dimensional cosmological manifolds whose spatial sub-manifolds have a `non-closed'  topology (in the sense defined in~\cite{Kleban:2016sqm}), will have slices of ever-growing volume. Moreover, these slices will contain a region where the expansion rate is faster than the one of de Sitter space in FRW slicing with the same cosmological constant. This is true even if locally singularities will form, such as black holes. It is quite remarkable that one can obtain general results with limited assumptions on the initial conditions and independently of the formation of singularities, where General Relativity (GR) breaks down.
These results, together with the regularity of MCF, strongly suggest that the 3-volume will go to infinity and that the regions that keep expanding will become locally indistinguishable from de Sitter space. In this paper we prove this statement in the simpler $2+1$-dimensional cosmology, postponing the---considerably more complicated---$3+1$ case to a future publication \cite{inprogress}. Since some of the results hold in any number of dimensions, we will keep the discussion general at first, and specify to $2+1$ dimensions only later.

Consider a cosmological spacetime, which is a $(n+1)$-dimensional manifold that can be foliated by $n$-dimensional Cauchy spacelike slices, $M_t$. These slices have all the same topology~\cite{Geroch}. A timeslice $M_{t}$ has induced metric $h_{\mu\nu}=g_{\mu\nu}+n_\mu n_\nu$, where $g_{\mu\nu}$ is the spacetime metric (we use the mostly-plus convention) and $n_\mu$ is orthonormal to $M_t$, $n_{\mu}n^{\mu} = -1$, and future-directed. The extrinsic curvature of these slices is defined as $K_{\mu\nu} \equiv h_\mu^{\;\alpha} \nabla_\alpha n_\nu$, satisfying $n^{\mu} K_{\mu \nu} = 0$ and with trace $K \equiv h^{\mu\nu}K_{\mu\nu} = g^{\mu \nu} K_{\mu \nu}$, and traceless part $\sigma_{\mu \nu} \equiv K_{\mu \nu} - {1 \over n} K h_{\mu \nu}$. (With our sign convention $K>0$ corresponds to expansion.) We also define $\sigma^2 \equiv \sigma_{\mu\nu} \sigma^{\mu\nu}$; notice that $\sigma^2 \geq 0$, since $\sigma_{\mu\nu}$ is a tensor projected on the spatial hypersurfaces.  

We will use the MCF of codimension-one spacelike surfaces in Lorentzian manifolds. This is defined as the deformation of a slice as follows: $y^\mu(x,\lambda)$ is,  at each $\lambda$, a mapping between the initial spatial manifold ${{M}}_0$, (which is parametrized by $x$) and the global spacetime, ${{M}}_0 \times [0,\lambda_0) \to M_{n+1}$. The evolution under the change of $\lambda$ is given by (see for instance \cite{EH})
\be
\frac{d}{d\lambda}y^\mu(x,\lambda)=K n^\mu (y^\alpha)\ ,
\ee 
where $n^\mu$ is the future-oriented vector orthonormal to the surface of constant $\lambda$.  

\begin{figure}
\begin{center}
\includegraphics[width=11cm,draft=false]{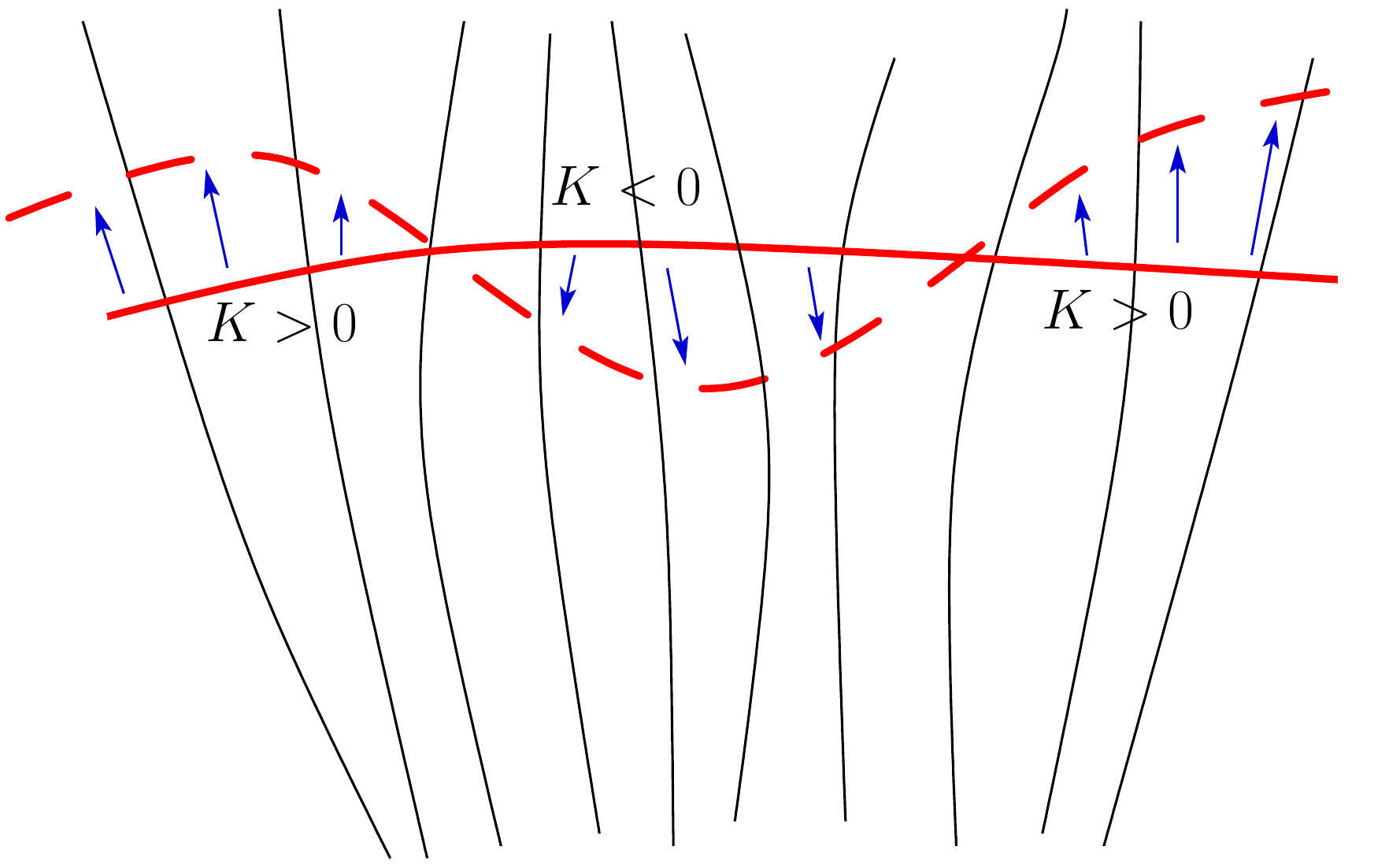}
\end{center}
\caption{\small
Pictorial depiction of Mean Curvature Flow. The new surface has larger or equal volume than the previous one. Figure from~\cite{Kleban:2016sqm}.  \label{fig:MKF} }
\end{figure}

Using the first variation of area formula
  \be\label{normdeform}
{\cal{L}}_{n} \log\sqrt{h}=K\ ,
 \ee
one gets the variation of $\sqrt{h}$ under the flow: ${d \over d \lambda} \sqrt{h} = K^2 \sqrt{h}$.  Therefore  the total spatial volume
$V \equiv \int_{M_t} d^n x \sqrt{h}$ satisfies
\be \label{volgrowth}
{d V \over d \lambda} = \int d^n x \sqrt{h}\,K^2 \geq 0 \;,
\ee
where $\lambda$ is the affine parameter of the deformation. Hence after the deformation, the new surface has either strictly larger or equal volume (see Fig.~\ref{fig:MKF}).  MCF has been very much studied in the context of Riemannian manifolds, but there is quite a large literature also for the Lorentzian (or semi-Riemannian) one, see~\cite{gerhardtbook}. 

The use of MCF is particularly useful in  our context because of the following two properties. First, in the Lorentzian case, this flow is endowed by many regularity  properties~\cite{gerhardtbook}, as it is quite intuitive from the fact that the maximization of the volume stretches the surface, making it smoother. Second, when spacelike crushing spacetime singularities~\cite{Eardley} form, the mean curvature flow remains at a finite distance from them~\cite{Kleban:2016sqm}. Again, this follows intuitively from the fact that the MCF tends to maximize the spatial volume. Therefore, MCF is a mathematical tool that allows us to explore the spacetime manifold without assuming the absence of crushing singularities~\footnote{Technically, however, we need to assume, as we will do in this paper, the absence of finite-volume singularities, {\it i.e.}~that the expanding spacetime simply comes to an end on some spacelike surface that is not a zero-volume crunch (see~\cite{Kleban:2016sqm}). Such finite-volume singularities seem to us highly artificial and unphysical on general grounds, and can probably be forbidden with appropriate conditions on the stress-energy tensor, see~\cite{barrow1986closed}.}. 

While the results of~\cite{East:2015ggf,Kleban:2016sqm} might be enough to dispense of the inflationary initial-patch problem, they convincingly suggest a stronger result: in the presence of a positive cosmological constant, for the `non-closed' topologies and with some mild conditions on the matter stress-tensor, initially everywhere-expanding manifolds will {\it always} become asymptotically indistinguishable from de Sitter space almost everywhere, {\it independently} of the initial conditions. 

Let us give some intuition of why this should be true. Since, from~\cite{Kleban:2016sqm}, some regions must keep expanding with a fast expansion rate, we expect that some region of the MCF-surface will grow in volume, while some others, in front of a crushing singularity, will stop evolving. As the volume grows, inhomogeneities and the matter density dilute away, the cosmological constant dominates, and the expanding regions become locally indistinguishable from de Sitter space. The MCF slices will therefore reach infinite volume, while only some regions, with vanishing relative volume, will not converge to de Sitter space (hence the specification `almost everywhere'). 

We provide here the proof of this statement in 2+1 dimensions, where the topology of the spatial manifolds has stronger implications on their geometry. Even though this is physically of limited interest, the theorem we will obtain is rather compelling and many of the results we will discuss are useful for the 3+1 case, the case of ultimate interest \cite{inprogress}. 

{\bf Notation and conventions.} The Riemann tensor is defined through $(\nabla_\mu\nabla_\nu - \nabla_\nu\nabla_\mu) \omega_\rho = R_{\mu\nu\rho}^{\quad\;\sigma} \omega_\sigma$, the Ricci tensor by $R_{\mu\nu} \equiv R_{\mu\sigma\nu}^{\quad\;\sigma}$, the Ricci scalar by $R \equiv R_\mu^{\;\mu}$. The Einstein equations are given 
\be
R_{\mu\nu}-\frac12 g_{\mu\nu} R = 8 \pi G (T_{\mu\nu}-\Lambda g_{\mu\nu}) \;,
\ee
where $\Lambda$ is the cosmological constant and $T_{\mu\nu}$ is the stress-energy tensor of all the other forms of matter.
The Ricci scalar associated with the induced metric $h_{\mu\nu}$ on the $n$-dimensional slices is denoted by $^{(n)}R$.

\section{\label{sec:theorem}Asymptotic behavior of 2+1 dimensional manifolds}

We will prove a theorem that requires the following assumptions:

\begin{itemize}

\item There is  a ``cosmology'', which is defined as a connected $n+1$ dimensional spacetime $M_{n+1}$ with a compact Cauchy surface.  This implies that the spacetime is topologically  $R \times M$ where $M$ is a compact $n$-manifold, and that it can be foliated by a family of topologically identical Cauchy surfaces $M_t$ \cite{Geroch}. We fix one such foliation,  {\it i.e.}~such a time function $t$, with $t\in[t_0,+\infty)$, and with associated lapse function $N$: $N^{-2} \equiv -\partial_\mu t \partial^\mu t$, $N >0$. We consider manifolds that are initially expanding everywhere, {\it i.e.}~there is an initial slice,  $M_{t_0}$, where $K>0$ everywhere (for example this holds if one has a global crushing singularity in the past).

\item There is a positive cosmological constant and matter that satisfies the Dominant Energy Condition (DEC) and the Strong Energy Condition (SEC). The DEC states that $-T^\mu{}_\nu k^\nu$ is a future-directed timelike or null vector for any future-directed timelike vector $k^\mu$. The DEC implies the Weak Energy Condition (WEC),  $T_{\mu \nu}k^{\mu}k^{\nu} \geq 0$ for all time-like vectors~$k^{\mu}$. The SEC, in $n+1$ dimensions, reads: $(T_{\mu\nu}- \frac{1}{n-1}g_{\mu\nu} T) k^\mu k^\nu  \geq 0$ for any future-directed timelike vector $k^\mu$.

\item  We will also need a technical assumption, see Definition~\ref{def:crushing}: the only spacetime singularities are of the crushing kind~\cite{Eardley} (thus singularities that have zero spatial volume). Physically, these are the only singularities that are believed to be relevant.

\item Our main result is for the case $n=2$. In this case, the  topology of the spatial 2-manifolds must not be ``closed'', meaning that the Euler characteristic is non-positive (in the case of the sphere, one additional  assumption is required, as we discuss later).

\end{itemize}

Let us comment on the physical restrictions implied by the above hypotheses. The SEC and the DEC are satisfied by non-relativistic matter, radiation and the gradient energy of a scalar field \footnote{For SEC indeed
\be
T_{\mu\nu} = \partial_\mu\phi \partial_\nu\phi - \frac12 g_{\mu\nu} (\partial\phi)^2 
\quad\Rightarrow\quad  \left(T_{\mu\nu}- \frac{g_{\mu\nu}}{n-1} T\right) k^\mu k^\nu = (\partial \phi \cdot k)^2 \geq 0\;.
\ee
}. The inflationary potential violates SEC and if the potential is negative somewhere also DEC is violated. However, in our setup the inflationary potential is represented by the positive cosmological constant, which is a good approximation in the inflationary region of the potential. 

We also comment on the definition of a crushing singularity, as we adopt a slight generalization of the Definitions 2.10 and 2.11 in~\cite{Eardley}. Our definition will agree with theirs in the case of asymptotically flat spacetimes.

\begin{definition}\label{def:crushing} Analogously to Definition 2.9 of~\cite{Eardley}, a future crushing function $\tilde t$ is a globally defined function on $M_{n+1}$ such that on a globally hyperbolic neighborhood ${\cal{N}}\cap\{\tilde t>c_0\}$, $\tilde t$ is a Cauchy time function with range $c_0<\tilde t<+\infty$ ($c_0 \ge 0$ is a constant), and such that the level sets~$S_c=\{\tilde t=c\}$, with $c>c_0$, have mean curvature $\tilde K<-c$.~\footnote{For example in a Schwarzschild-de Sitter spacetime in the standard coordinates, one could take $\tilde t$ to be a function of $r$ for $r$ close to 0, so the level sets $S_c$ would be $r={\rm const}$.}  We shall say that a Cosmology has potential singularities only of the crushing kind if there is an open set ${\cal{N}}$ such that, outside ${\cal{N}}$, the inverse of the lapse of the $t$ foliation, $N^{-1}$, is bounded, and such that ${\cal{N}}$ contains a Cauchy slice and admits a future crushing function $\tilde t$ and, for any given $c$, in $\{\tilde t \leq c\}$, $N^{-1}$ is bounded. 
\end{definition}

\noindent In physical terms,  this ${\cal{N}}$ corresponds to a subset of the interior of black holes, and we are requiring that any possible pathology takes place only for $\tilde t\to \infty$. \\

Under these assumptions, we will prove the following:

\begin{theorem}\label{theorem:main}
The spatial volume goes to infinity. At late times, {\it i.e.}~for large values of the MCF affine parameter $\lambda$, the spacetime converges, on average, to de Sitter space; and there are arbitrarily large regions of space-time that are physically indistinguishable from de Sitter space.
\end{theorem}

\subsubsection*{Proof of the theorem}

{\bf Existence of the flow:} We first establish the existence of MCF for arbitrary flow parameter $\lambda$. 

The MCF evolution of $K$ (see for example \cite{EH}) is given by~\footnote{One can get this equation similarly to the derivation of the Raychauduri equation, bearing in mind that we are {\em not} following geodesics orthogonal to the surface. 
\be\label{derKdiff}
\begin{split}
\frac{d K}{d \lambda} = &K n^\alpha \nabla_\alpha(\nabla_\mu n^\mu) = K n^\alpha \nabla_\mu\nabla_\alpha n^\mu - R_{\alpha\mu\rho}^{\quad\;\mu} n^\rho n^\alpha K = K \nabla_\mu(n^\alpha \nabla_\alpha n^\mu) - K \nabla_\mu n^\alpha \nabla_\alpha n^\mu - R_{\alpha\rho}n^\alpha n^\rho K = \\
= & K \nabla_\mu(n^\alpha \nabla_\alpha n^\mu) - K K_{\mu\nu} K^{\mu\nu} - R_{\alpha\rho}n^\alpha n^\rho K \;.
\end{split}
\ee
Imposing that $n^\mu$ remain perpendicular to the surface one gets
\be
K n^\alpha \nabla_\alpha n^\mu = h^{\mu\nu} \nabla_\nu K \;.
\ee
Therefore
\be
K \nabla_\mu(n^\alpha \nabla_\alpha n^\mu) = \nabla_\mu (h^{\mu\nu} \nabla_\nu K) - \nabla_\mu K n^\alpha \nabla_\alpha n^\mu = \Delta K - \nabla_\mu(h^{\alpha\nu} \nabla_\nu K) n^\mu n_\alpha  - \nabla_\mu K n^\alpha \nabla_\alpha n^\mu = \Delta K \;,
\ee
where in the last step, to obtain the Laplacian on the surface, we separated the covariant derivative in the components parallel and orthogonal to the surface. Plugging this expression in eq.~\eqref{derKdiff} one obtains eq.~\eqref{eq:MCF} after separating the contribution of the cosmological constant and the traceless part of $K_{\mu\nu}$.
}
\be\label{eq:MCF}
\frac{d K}{d \lambda} -\Delta K + \frac{1}{n} K\left(K^2-K_\Lambda^2\right) + \sigma^2 K + R^{(m)}_{\mu\nu} n^\mu n^\nu K =0 \;,
\ee
where $\Delta$ is the Laplacian operator on the surface, $K_\Lambda^2 \equiv \frac{n}{n-1}16\pi G \Lambda>0$, and
\be
R^{(m)}_{\mu\nu} \equiv 8\pi G \left(T_{\mu\nu}-\frac{g_{\mu\nu}}{n-1} T\right) \ .
\ee
The SEC gives
\be\label{eq:SEC}
R^{(m)}_{\mu\nu} n^\mu n^\nu \geq 0 \;.
\ee
It is worthwhile to mention two properties of the evolution under MCF. First, if a surface is spacelike, it remains so: in fact the local volume form is non-decreasing under MCF, but it would vanish if the surface became null anywhere~(see for example~\cite{Kleban:2016sqm}).  Second, it also preserves the property that $K>0$ everywhere  (see \emph{e.g.} \cite{gerhardtbook}, Proposition 2.7.1).  Intuitively, this is because the flow stops in any region where $K$ approaches zero.

Since $M_\lambda$ is compact, $K(x,\lambda)$ has a maximum at each $\lambda$, $K_m(\lambda)\equiv  {\rm Max}_{x} K(x, \lambda)$. Let us observe the evolution of this point. Intuitively, if  $K_m(\lambda)\geq K_\Lambda$, then all terms of (\ref{eq:MCF}) but the first are non-negative (around a maximum the Laplacian is non-positive), which implies that the maximum is non-increasing with affine time, as long as it is larger than $K_\Lambda$. This suggests that $K_m(\lambda)$ is bounded by a quantity that goes to $K_\Lambda$ as $\lambda\to \infty$. Indeed this can be proven rigorously, and we can also bound the rate of convergence:

\begin{theorem}\label{th:boundonmax}
Let $M_\lambda$ be smooth compact spacelike hypersurfaces satisfying the MCF equations, in an interval $[\lambda_1,\lambda_2]$, inside the smooth $(n + 1)$-dimensional Lorentzian manifold  $M_{n+1}$. Suppose also there exists a point $(x,\lambda)$, with $\lambda_1\leq\lambda\leq\lambda_2$, such that $K(x,\lambda)>K_\Lambda$, then  we have
\be\label{eq:maxbound}
K_m(\lambda_2)\leq K_\Lambda+ e^{- \frac{2}{n}K_\Lambda^2(\lambda_2-\lambda_1)}(K_m(\lambda_1)-K_\Lambda) \ ,
\ee
so the maximum, if larger than $K_\Lambda$, decays exponentially fast towards $K_\Lambda$ with a rate given by the cosmological constant.
\end{theorem}

Notice that if no  point $(x,\lambda)$ as in the hypotheses of the theorem exists, then the maximum $K_m(\lambda)$, with $\lambda_1\leq\lambda\leq\lambda_2$, is automatically $\leq K_\Lambda$.\\

{\bf Proof}: For an $(n + 1)$-dimensional Lorentzian manifold, eq.~(\ref{eq:MCF}) can be put in the following form
\be
\partial_\lambda (K-K_\Lambda)-\Delta(K-K_\Lambda)+\frac{1}{n}K(K+K_\Lambda)(K-K_\Lambda)+\text{non-negative terms} =0 \;.
\ee
Multiply by $e^{\alpha\lambda}$, where $\alpha$ is a real constant, to obtain
\bea\nn
&&\partial_\lambda (e^{\alpha\lambda} (K-K_\Lambda))-\Delta (e^{\alpha\lambda}(K-K_\Lambda))
+\frac{1}{n}\left(K(K+K_\Lambda)-n\,\alpha\right) e^{\alpha\lambda}(K-K_\Lambda)\\ 
&&+\text{non-negative terms} =0  \ .
\eea
Let us think of this as an equation for $W\equiv e^{\alpha\lambda} (K-K_\Lambda)$. Consider the compact interval $[\lambda_1,\lambda_2]\times M,$ and notice that, by assumption,  $W$ is positive at one point here. Then, since $W$ is continuous and the interval is compact, the maximum of $W$ is attained, and this is $>0$ as, by hypothesis, $K>K_\Lambda$ at one point. At the maximum, if this is not at $\lambda=\lambda_1$, the first term of the equation is $\geq 0$ (it is $0$ if the maximum is in the interior of the interval and $\geq 0$ if it is in $\lambda_2$), the second $\geq 0$ as well (with the $-$ sign included), the third $>0$ provided $0\leq \alpha\leq{\frac{2}{n}} K_\Lambda^2$ (since $K>K_\Lambda$ at the maximum), the rest $\geq 0$, so we obtain $0>0$, a contradiction. The conclusion is that the maximum is attained at $\lambda=\lambda_1$, and so
\be\label{eq:maxbound2}
K_m(\lambda_2)-K_\Lambda\leq e^{-\alpha(\lambda_2-\lambda_1)}(K_m(\lambda_1)-K_\Lambda)\ ,
\ee
which, by choosing $\alpha={\frac{2}{n}}K_\Lambda^2$, ends the proof of Theorem~\ref{th:boundonmax}.\\

 There is powerful theorem for the MCF of codimension-one spacelike surfaces in Lorentzian manifolds that guarantees the regularity of the flow and therefore the existence of the MCF as long as it is contained in a compact regular region of the spacetime manifold. In detail:

\begin{theorem}\label{theorem:existence}
~\cite{EH,E} Let $M_{n+1}$ be a smooth $(n + 1)$-dimensional Lorentzian manifold satisfying the WEC. Let $M_0$ be a compact smooth spacelike
hypersurface in $M_{n+1}$. Then there exists a unique family ($M_\lambda$) of smooth compact spacelike hypersurfaces satisfying the MCF equations, in an interval $[0,\lambda_0)$ for some $\lambda_0>  0$ and having initial data $M_0$. Moreover, if this family stays inside a smooth compact region of $M_{n+1}$ then the solution can be extended beyond $\lambda_0$.
\end{theorem}

We are now going to show that, given our hypotheses, we can apply this theorem and show that the flow exists for arbitrarily large $\lambda$. We will first show that as long as the flow stays sufficiently far from crushing singularities, {\em i.e.}~outside one of the level set of $\tilde t$, $S_c$, then, for bounded $\lambda$, it stays in a compact region (Theorem \ref{theorem:noinf}). Then we are going to show that there is a $c$ such that for bounded $\lambda$ (hence for all $\lambda$'s) the flow cannot meet $S_c$ (Theorem \ref{theorem:nocrushing}). Hence the hypotheses and therefore the conclusions of Theorem \ref{theorem:noinf} are always satisfied and  the flow exists globally for arbitrary $\lambda$.

One can always choose the global time function $t$ in such a way that $N^{-1}$ is bounded over the whole manifold, except potentially as we approach the crushing singularities. Since they can be reached in a finite proper time, $N$ will typically go to zero there. The unit vector perpendicular to the global time surfaces is $N \partial_\mu t$; it forms an angle with the unit vector perpendicular to the MCF surfaces $n^\mu$: $v \equiv N \partial_\mu t  \; n^\mu$, $v \geq 1$.\footnote{The crucial step in proving the existence theorem \ref{theorem:existence} is to show that $v$ cannot diverge in a compact region of the manifold \cite{EH,E}.}  As discussed above, the MCF gives at each $\lambda$ a mapping between the initial spatial manifold (parametrised by $x$) and the global spacetime, ${{M}}_0 \times [0,\lambda_0) \to M_{n+1}$.  Let $u(x,\lambda)$ be the value of $t$ at the image of $(x,\lambda)$ under the flow. To prove that the flow does not get to infinity at finite $\lambda$ means to have a bound on the growth of $u$ as a function of $\lambda$. This is given by the following theorem.


\begin{theorem}\label{theorem:noinf}
Let $c>c_0$, with $c_0$ given in Definition~\ref{def:crushing}. There exists a constant $C \geq 0$ (depending on the $\sup_{\tilde t\leq c} N^{-1}$ and $\sup_{M_0} K$) such that the following holds.

Let $\lambda_0>0$. Provided that $M_{\lambda}$ is in $\tilde t<c$ for all $\lambda\in[0,\lambda_0)$, we have
\be
u(x,\lambda)\leq \sup_x u(x,0) + C \lambda
\ee
for all $x$ and $\lambda\leq\lambda_0$.
\end{theorem}

We remark that if $M_\lambda$ is in $\tilde t<c$ for all $\lambda\geq 0$, then we can apply the theorem with any $\lambda_0>0$, with $C$ independent of $\lambda_0$, hence the estimate of the theorem holds for all $\lambda\geq 0$.
\\

{\bf Proof}:
Consider the function $U(x,\lambda)\equiv u(x,\lambda) - C \lambda$ in the compact set $M\times[0,\lambda_0]$. $U$ attains the max somewhere, say at $(x_1,\lambda_1)$. Now, for fixed $\lambda$, the max of $U$ is attained where the max of $u$ is attained, so $x_1$ is in fact one of the maximum locations of $u$ for parameter $\lambda_1$, and thus $v=1$ there. On the other hand,
\be
\frac{\partial U}{\partial\lambda}=K N^{-1} v-C \;.
\ee
In particular, at the maximum $\{x_1,\lambda_1\}$, we have
\be
\left.\frac{\partial U}{\partial\lambda}\right|_{\{x_1,\lambda_1\}}=K N^{-1} -C\ ,
\ee
and if this max is attained at $\lambda_1>0$, then
\be
\left.\frac{\partial U}{\partial\lambda}\right|_{\{x_1,\lambda_1\}}\geq 0 
\ee
(it is $0$ if the maximum is in the interior of the interval and $\geq 0$ if it is at $\lambda_0$). So, at the point $\{x_1,\lambda_1\}$, we must have
\be
C\leq K N^{-1}\ .
\ee
But for large enough $C$, this contradicts the previous bounds on $K$ (theorem \ref{th:boundonmax}) and on~$N^{-1}$. By choosing $C\geq \sup_{\{x,\lambda\}} K \cdot \sup_{\tilde t\leq c} N^{-1}$, the max of $U$ is attained at $\lambda_1=0$. As we will show later, on the initial surface $M_0$, $K_m(0)\geq K_\Lambda$ and therefore, by theorem \ref{th:boundonmax}, $\sup_{\{x,\lambda\}} K\leq \sup_{\{x,\lambda=0\}} K$. The claim follows. \\

We now show the rather intuitive fact that the flow stays away from the crushing singularities, and therefore, given our hypothesis on the nature of the singularities, stays away from these (see also~\cite{Kleban:2016sqm,Mirbabayi:2018suw}):

\begin{theorem}\label{theorem:nocrushing}
If the initial surface of the flow has $K\geq 0$, then $M_\lambda$ stays away from a crushing singularity.

More precisely, for any $c>c_0>0$ such that $M_0$ is in $\{\tilde t<c\}$, the flow remains in $\{\tilde t<c\}$.
\end{theorem}

\begin{figure}[t!] 
\centering
{\includegraphics[scale=.60,angle=0]{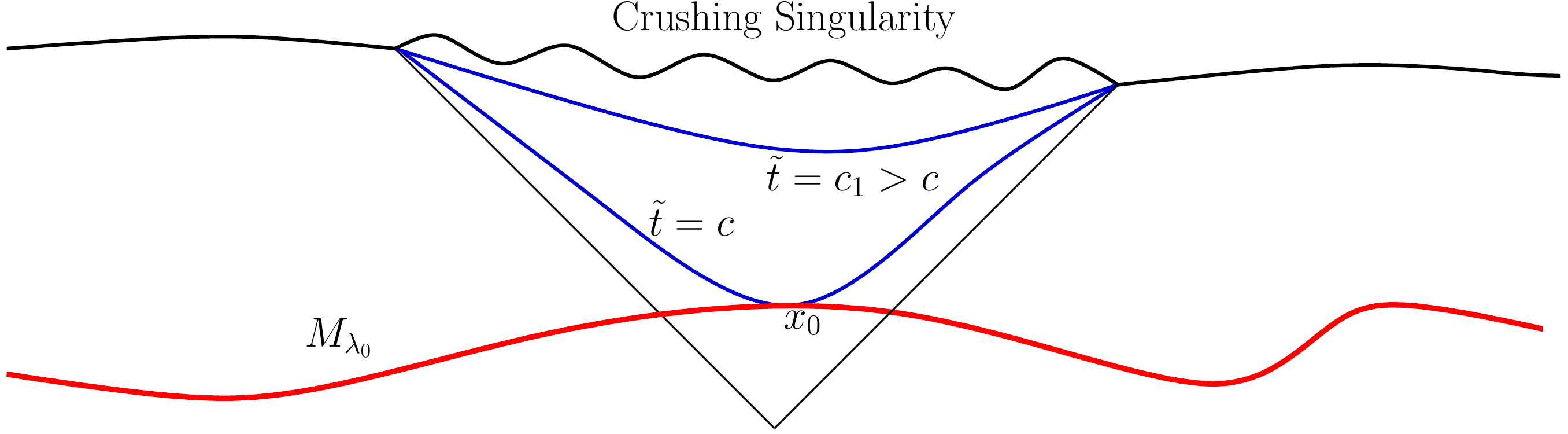}} \quad\quad\quad
\caption{\small Geometry of theorem 5. The surface of MCF at the hypothetical $\lambda_0$ where it becomes tangent to a level-surface $S_{c}$ of $\tilde t=c>0$. $\tilde t$ is the time function associated to the crushing singularity. }\label{fig:nocrushing}
\end{figure} 

{\bf Proof:} Suppose, for the sake of contradiction, that there exists an affine parameter $\lambda>0$ at which the MCF surface intersects the level-surface $S_c$ of $\tilde t$ (see Fig.~\ref{fig:nocrushing}); let $\lambda_0$ be the infimum of the set of these values of $\lambda$.

We claim that $M_{\lambda_0}\cap S_c$ is non-empty: if $\lambda_j>\lambda_0$ with $\lim_j \lambda_j=\lambda_0$ and $M_{\lambda_j}\cap S_c$ non-empty (which exists by the definition of $\lambda_0$), let $x_j\in M_{\lambda_j}\cap S_c$ and let $y_j$ be the corresponding point in the initial surface $M_0$; by the compactness of $M_0$ a subsequence converges to some $y_0\in M_0$, and then by the continuity of the MCF, the image of $y_0$ at affine parameter $\lambda_0$ is $x_0$, the limit, along the subsequence, of the $x_j$, which is thus in $S_c\cap M_{\lambda_0}$ (as $S_c$ is closed). At $\lambda_0$ the surface is actually tangent to $S_c$ at the point $x_0$ since necessarily $x_0$ is a maximum of $\tilde t$ along $M_{\lambda_0}$, so that $d\tilde t$ is conormal to $M_{\lambda_0}$ at $x_0$, as it is to $S_c$. Moreover, $S_c$ has, by definition~\ref{def:crushing}, extrinsic curvature everywhere $<-c<0$. 

Bartnik~\cite{bartnik1984} gives us a useful expression for the Laplacian of the restriction, $\tilde u$, of a time-like function,~$\tilde t$, to a spacelike surface, $\tilde S$. We have (see eq.~(2) of~\cite{EH}):
\be
\Delta \tilde u= K_{\tilde S} \tilde N^{-1} \tilde v+{\rm div}\bar{\nabla} \tilde t\ ,
\ee
where $K_{\tilde S}$ is the extrinsic curvature of the surface, $\tilde N$ is the lapse in the $\tilde t$-coordinates, $\tilde v$ is the angle between the normal vector and the gradient of~$\tilde t$, $\bar \nabla$ is the gradient in the ambient space, and ${\rm div}$ of a vector is the trace of the projection on the surface of the covariant derivative of that vector\footnote{This formula can be easily derived. Consider the gradient of the height function projected on the surface 
\be
\nabla_\parallel{}^\mu \tilde u = \nabla_\parallel{}^\mu \tilde t = (g^{\mu\nu} +n^\mu n^\nu) \partial_\nu \tilde t = \partial^\mu \tilde t + \tilde v \tilde N^{-1} n^\mu \;.
\ee
We can now take a second derivative and trace it on the surface
\be
\Delta \tilde u =  \nabla_{\parallel}{}_{\mu} \nabla_\parallel{}^\mu \tilde u= h^\nu_\mu \bar\nabla_\nu \nabla_\parallel{}^\mu \tilde u =h^\nu_\mu \bar\nabla_\nu \partial^\mu \tilde t + h^\nu_\mu \bar\nabla_\nu (\tilde v \tilde N^{-1} n^\mu) = {\rm div}\bar{\nabla} \tilde t + \tilde v \tilde N^{-1} K \;,
\ee
where
\be
{\rm div}\bar{\nabla} \tilde t \equiv h^\nu_\mu \bar\nabla_\nu \partial^\mu \tilde t \;.
\ee
}. We are going to apply this formula to the two surfaces that are tangent to each other at $(\lambda_0,x_0)$: the MCF one at parameter $\lambda_0$, and a level-set of~$\tilde t$,~$S_c$ (in this case $\tilde u$ is a constant).

When we apply it to the MCF surface at parameter $\lambda_0$, at the tangent point $x_0$, {\em i.e.}~$M_{\lambda_0}\cap S_c$, we have
\be\label{eq:bartnik1}
\Delta\tilde u= K \tilde N^{-1}\tilde v+{\rm div}\bar{\nabla}\tilde t\ .
\ee
When we apply it to $S_c$ at the same point, we have
\be\label{eq:bartnik2}
0=\tilde K_c\tilde N^{-1}\tilde v+{\rm div}\bar{\nabla}\tilde t \ ,
\ee
where $\tilde K_c$ is the extrinsic curvature of $S_c$, and where we used that the Laplacian on the surface vanishes as the surface is at constant $\tilde t$. Notice that at this point the two surfaces are tangent and therefore the normal vectors of the two surfaces are the same.
Taking the difference between (\ref{eq:bartnik1}) and (\ref{eq:bartnik2}), we have
\be\label{eq:bartnik3}
\Delta\tilde u=(K-\tilde K_c)\tilde N^{-1}\tilde v \;.
\ee
For fixed $\lambda_0$, at the tangent point $x_0$, if it existed, the height would reach a maximum, which would imply $\Delta\tilde u\leq 0$. However, $K-\tilde K_c>0$, as $K\geq 0$, given that MCF preserves the property that $K\geq 0$ (see \emph{e.g.} \cite{gerhardtbook}, Proposition 2.7.1), and $\tilde K_c<-c<0$ by definition of $S_c$. Therefore the left-hand side of~(\ref{eq:bartnik3}) is non-positive, while the right-hand side is positive. We have reached a contradiction, and the theorem is proved. \\

Notice, as a corollary, one can prove the stronger statement that the MCF cannot cross {\em any} level set with $c>c_0$. This follows from the argument above, if one proves that that initial surface cannot cross the level set. This again can be shown by contradiction. 
If the initial surface crosses level sets with $c>c_0$ one can take the maximum $c$ for which this happens. This level set is now tangent to the initial surface and one runs into contradiction using the same argument above. 
\\

For any $\lambda_0>0$ such that the flow is defined for affine parameter $\lambda\in[0,\lambda_0)$, $\tilde t$ remains bounded above by Theorem~\ref{theorem:nocrushing}, so by Theorem~\ref{theorem:noinf} the height function is bounded by an affine function of $\lambda$. Hence, the MCF remains in a compact region of the manifold for $\lambda<\lambda_0$: by theorem \ref{theorem:existence} the MCF can be extended beyond $\lambda_0$, and thus exists for arbitrarily large $\lambda$.  \\

As a side comment, we finally notice that  the existence of the flow and the fact that the flow stays away from crushing singularities, {\it i.e.} Theorems \ref{th:boundonmax},\ref{theorem:existence},\ref{theorem:noinf},\ref{theorem:nocrushing}, remain true even if, instead of a positive cosmological constant and matter satisfying SEC, one has a non-constant scalar-field potential (which violates SEC), as long as the potential energy has positive upper and lower bounds $\Lambda_2$ and $\Lambda_1$. In fact, Theorem~\ref{th:boundonmax} holds by substituting $K_\Lambda\to K_{\Lambda_2} \equiv \sqrt{\frac{n}{n-1}16\pi G \Lambda_2}$ and noticing that a negative potential satisfies SEC~\footnote{In fact, one can write eq.~(\ref{eq:MCF}) as 
\be\label{eq:MCF2}\nn
\frac{d K}{d \lambda} -\Delta K + \frac{1}{n} K\left(K^2-K_{\Lambda_2}^2\right) + \sigma^2 K + \bar R^{(m)}_{\mu\nu} n^\mu n^\nu K =0 \;,
\ee
with 
\be\nn
\bar R^{(m)}_{\mu\nu} \equiv 8\pi G \left(T_{\mu\nu}-\frac{g_{\mu\nu}}{n-1} T\right)+\frac{16\pi G}{n-1} g_{\mu\nu}\left(V(\phi)-\Lambda_2\right)\ ,
\ee
where $T_{\mu\nu}$ is the part of the stress tensor that  satisfies SEC. Given the upper bound on $V(\phi)$, $\bar R^{(m)}_{\mu\nu} n^\mu n^\nu\geq0$, and Theorem~\ref{th:boundonmax} follows.}. Similarly, Theorem~\ref{theorem:noinf} holds by replacing $\sup_{M_0} K$ with $\sup_{\{x,\lambda\}} K$, as in this case, by Theorem~\ref{th:boundonmax}, $K$ has a bounded sup on the flow, but it is not guaranteed that this is attained on the initial surface. These observations represent the starting point for a possible generalization of the results that we are going to discuss next to the case of a full-fledged slow-roll inflationary model.\\

So far, our results are valid in any number of dimensions, $n\geq 2$. From now on, we specify to the particular case of 2+1 dimensions. \\

{\bf Infinite Volume:} Now that we have proven that the flow exists for all $\lambda$'s, we are going to prove that the manifold reaches infinite volume. We will do so by proving that $M_\lambda$ reaches infinite volume.
We start by noticing that, by using the Gauss-Codazzi relation (see for instance \cite{Wald84}, eq.~(E.2.27)), Einstein's equations, contracted with $n^\mu n^\nu$ give:
\be
\label{H}
{^{(2)}\!R} + \frac 12 K^2 - \sigma^2 = \frac12 K_\Lambda^2 + 16\pi G T_{\mu\nu} n^\mu n^\nu \;.
\ee

We consider an initial surface that is expanding, $K\geq 0$ everywhere. Under MCF the volume evolves as
\be
\begin{split}
\label{MCF}
\frac{d V}{d \lambda} & = \int d^2 x \sqrt{h} \; K^2 = \int  d^2x \sqrt{h} \left(32 \pi G T_{\mu\nu}n^\mu n^\nu + K_\Lambda^2+ 2 \sigma^2 - 2 \cdot {^{(2)}\!R}\right) \ge \\ &\geq  K_\Lambda^2\int d^2 x \sqrt{h} - 2 \int d^2 x \sqrt{h} {^{(2)}\!R} = K_\Lambda^2 V - 8 \pi \chi \;.
\end{split}
\ee
Here we have used that, by  WEC, $T_{\mu\nu}n^\mu n^\nu \geq 0$. The Euler characteristic is positive only for the sphere, zero for the torus and negative for the rest. For $\chi \leq 0$ the volume goes to infinity. This establishes the first sentence of theorem~\ref{theorem:main}. In the case of the sphere ($\chi =2$) one has to compare the two terms at the initial conditions: if the $\Lambda$ term wins
\be\label{volumeS2}
V(0) >    \frac{16 \pi}{K_\Lambda^2}\ ,
\ee
 the volume goes to infinity.
 
 The explicit solution of (\ref{MCF}) reads 
 \be
 V(\lambda)\geq \frac{8 \pi\chi }{K_\Lambda^2}+ e^{K_\Lambda^2 \lambda}\left(V(0)-\frac{8 \pi\chi}{K_\Lambda^2}\right)\ .
 \ee 
 For all topologies except the sphere, $\chi \leq 0$, one has
 \be\label{eq:Vmin}
 V(\lambda)\geq V(0)\; e^{K_\Lambda^2 \lambda} \left(1 - \frac{8 \pi \chi}{K_\Lambda^2 V(0)} + {\cal {O}}(e^{-K_\Lambda^2\lambda})\right)\to\  \infty\ ,
 \ee 
 independently of the size of the initial volume.
 
 Notice that the rate at which $V$ goes to infinity is larger or equal than the one of the FRW slices of de Sitter space. We now prove that, at late times, the volume goes to infinity with the same rate as these slices of de Sitter. In fact, for manifolds where the Euler characteristic $\chi$ is non-positive,  there must be a point where ${^{(2)}\!R} < 0$. At that point, since $\sigma^2\geq0$ and $T_{\mu\nu} n^\mu n^\nu\geq0$ (by the WEC), eq.~(\ref{H}) implies that $K > K_\Lambda$ at that point (this is the argument of~\cite{Kleban:2016sqm} in 2+1 dimensions). The hypotheses of Theorem~\ref{th:boundonmax} therefore always hold in our case~\footnote{In the case of the torus one can have ${^{(2)}\!R} = 0$ everywhere, which allows $K \leq K_\Lambda$ everywhere. {Still, all our conclusions hold, as can be seen from the sentence below the formulation of Theorem~\ref{th:boundonmax}.  Notice, however, that this is a particularly simple case:} eq.~\eqref{H} would imply $K = K_\Lambda$ and $\sigma_{\mu\nu} =0$. Also $T_{\mu\nu}n^\nu n^\mu=0$ and this would imply $T_{\mu\nu} =0$ because of the DEC (see the discussion below). Therefore, one has the flat slicing of de Sitter space.
 }. This implies that eq.~(\ref{eq:maxbound}) bounds the maximum to be $K_\Lambda$ up to an exponentially small quantity. For the volume, we therefore have
 \bea\nn
&& \frac{d V}{d \lambda} =\int d^2 x \sqrt{h} \; K^2 \leq  \int d^2 x \sqrt{h} \; K_m(\lambda)^2\leq  \int d^2 x \sqrt{h} \;\left(K_\Lambda+ \left(K_m(0)-K_\Lambda\right) e^{-K_\Lambda^2\lambda}\right)^2\\ 
&&\qquad=\left(K_\Lambda+ \left(K_m(0)-K_\Lambda\right) e^{-K_\Lambda^2\lambda}\right)^2 V(\lambda)\ .
\eea
Thus
  \bea\label{eq:Vmax}\nn
&& V(\lambda)\leq V(0)\cdot e^{\left(\frac{1}{2}\left(\frac{K_m(0)}{K_\Lambda}+1\right)^2-2\right)} \cdot e^{K_\Lambda^2\lambda}\cdot  
    e^{ - \left(2\left(  \frac{K_m(0)}{K_\Lambda}-1\right) e^{-   K_\Lambda^2\lambda}+\frac{1}{2}\left(\frac{K_m(0)}{K_\Lambda}-1\right)^2 e^{-2   K_\Lambda^2\lambda}\right)}\\
   &&\qquad\to \quad V(0)\cdot e^{\left(\frac{1}{2}\left(\frac{K_m(0)}{K_\Lambda}+1\right)^2-2\right)} \cdot e^{K_\Lambda^2\lambda}\left(1+{\cal {O}}(e^{-K_\Lambda^2\lambda})\right) \;.
  \eea

  Combining with (\ref{eq:Vmin}), and keeping track of the signs of the subleading corrections in both equations, we obtain an expression for the volume that is valid at all times:
  \bea
&&  1  \leq \frac{V(\lambda)}{V(0) e^{K_\Lambda^2\lambda}}\leq e^{\left(\frac{1}{2}\left(\frac{K_m(0)}{K_\Lambda}+1\right)^2-2\right)}\ .
  \eea
Notice that by `restarting' the flow at sufficiently large $\lambda$, $K_m(0)$ gets arbitrarily close to $K_\Lambda$, so that the volume grows at late enough times as the FRW slicing of de Sitter space with arbitrary precision.\\

{\bf Stress Tensor:} We are now going to show that $T_{\mu\nu}$ becomes small in most of the volume. Eq.~\eqref{eq:maxbound} gives an upper bound on the quantity $ \int d^2x \sqrt{h} \; K^2$:
\be
\int d^2x \sqrt{h} \; K^2\leq K_\Lambda^2 V(\lambda)  \left(1+\left(\frac{K_m(0)}{K_\Lambda}-1\right) e^{-K_\Lambda^2\lambda}\right)^2  \;.
 \ee
 Let us now integrate (\ref{H}) over the MCF surface and use the inequality above:
 \bea\nn
4\pi\chi +\frac{1}{2}K_\Lambda^2 V(\lambda) \left(1+\left(\frac{K_m(0)}{K_\Lambda}-1\right) e^{-K_\Lambda^2\lambda}\right)^2-\int d^2x \sqrt{h} \;\sigma^2 \geq && \\ \geq \frac{1}{2} K_\Lambda^2 V(\lambda)+16\pi G  \int d^2x \sqrt{h} \; T_{\mu\nu} n^\mu n^\nu\;.  \label{eq:integrated}
\eea
This implies
\be\label{eq:sigmaTineq}
16\pi G \int d^2x \sqrt{h} \; T_{\mu\nu} n^\mu n^\nu +\int d^2x \sqrt{h}\; \sigma^2 \leq  c_1\ ,\\
 \ee
 where $c_1 $ is a non-negative constant. 
 Given that $\sigma^2\geq 0$ and $T_{\mu\nu} n^\mu n^\nu\geq 0$, this implies 
 \be\label{eq:sigmaTineq2}
16\pi G \int d^2x \sqrt{h} \; T_{\mu\nu} n^\mu n^\nu \leq c_1 \;,\qquad \int d^2x \sqrt{h}\; \sigma^2 \leq  c_1\ .\\
 \ee
Then $ \sigma^2$ and $G T_{\mu\nu} n^\mu n^\nu$ can be  $K_\Lambda^2 \cdot {\cal {O}}(1)$ at most for a physical volume that grows with $\lambda$ no more than a constant, and, similarly,  wherever $ \sigma^2$ and $G T_{\mu\nu} n^\mu n^\nu$ grow with $\lambda$, the associated physical volume must decrease accordingly. Therefore, in almost-all of the ever-growing volume, $ \sigma^2$ and $G T_{\mu\nu} n^\mu n^\nu$ have to be at most of order $K_\Lambda^2 \cdot {\cal {O}}(e^{-K_\Lambda^2\lambda})$.
 
 Because of the DEC, $T_{\mu\nu} n^\mu n^\nu$ is at least as large as the magnitude of any other component of the stress tensor in an orthonormal frame where $n^\mu$ is the timelike vector. We therefore define an associated vielbein $e_\mu{}^{a}$, such that $g_{\mu\nu}=e_\mu{}^{a}e_\nu{}^{b}\eta_{ab}$, with $\eta_{ab}$ being the Minkowski metric. We choose $e_\mu{}^0=n_\mu$. By DEC, we have
 \be\label{eq:Tmunubound}
  \int d^2x \sqrt{h}\;16\pi G\,  \left| T_{\mu\nu} e^{\mu a}e^{\nu b}\right|\leq  \int d^2x \sqrt{h}\;16\pi G\, T_{\mu\nu}n^\mu n^\nu \leq c_1\  .
 \ee
We therefore see that in almost-all of the ever-growing volume, $T_{\mu\nu}$ has to be at most of order~$K_\Lambda^2 \cdot {\cal {O}}(e^{-K_\Lambda^2\lambda})\to 0$. \\
 
 {\bf Ricci and Riemann Tensor:} We can now show that the Ricci tensor converges in almost all of the volume, to the one of de Sitter space. In fact, we can take the Einstein equations, contract them with $e^{\mu a}e^{\nu b}$ 
  \bea
&&  {R}_{\mu\nu} e^{\mu a}e^{\nu b} =\left[8\pi G \left(T_{\mu\nu}-T g_{\mu\nu}\right)+\frac{K_\Lambda^2}{2} g_{\mu\nu}\right] e^{\mu a}e^{\nu b}\ .
 \eea
 Let us write $R_{\mu\nu}$ as $R_{\mu\nu}=R_{dS,\mu\nu}+\delta R_{\mu\nu}$, where $R_{dS,\mu\nu}=\frac{ K_\Lambda^2}{2} g_{\mu\nu}$  is the Ricci tensor of de Sitter space with cosmological constant $\Lambda$. We obtain
 \bea
&&  {\delta R}_{\mu\nu}e^{\mu a}e^{\nu b} = \;8\pi G \left(T_{\mu\nu}-T g_{\mu\nu}\right)e^{\mu a}e^{\nu b} \ .
 \eea
We can now use the bound (\ref{eq:Tmunubound})
  \bea
&& \int d^2x\sqrt{h} \;\left|  {\delta R}_{\mu\nu}e^{\mu a}e^{\nu b}\right| =\int d^2x\sqrt{h} \;8\pi G \,\left| T_{\mu\nu}e^{\mu a}e^{\nu b}-T \eta^{ab}\right|\leq c_1  \ .
 \eea
 Therefore, as for $T_{\mu\nu}$, ${\delta R}_{\mu\nu}$ can be non-exponentially-vanishing  at most for a physical volume that grows with $\lambda$ no more than a constant. In almost-all of the ever-growing volume, ${\delta R}_{\mu\nu}$ has to be at most of order $K_\Lambda^2 \cdot {\cal {O}}(e^{-K_\Lambda^2\lambda})$.

We are now ready to show that the Riemann tensor tends to the one of de Sitter in the same sense as the Ricci tensor. This is so because, in 2+1 dimensions, the two are proportional to each other:
\be
R_{\mu\nu\rho\sigma}=g_{\mu\rho}\,R_{\nu\sigma}+g_{\nu\sigma}\,R_{\mu\rho}-g_{\mu\sigma}\,R_{\nu\rho}-g_{\nu\rho}\,R_{\mu\sigma}-\frac{1}{2}\left(g_{\mu\rho} g_{\nu\sigma}-g_{\mu\sigma}g_{\nu\rho}\right)\,R\ .
\ee
We therefore conclude that, apart for an infinitesimally small fraction, whose physical volume is at-most-finite, in all the rest of the ever-growing volume, the Riemann tensor becomes arbitrarily close to the one of de Sitter. Once averaged over the volume, this implies that on average the Riemann tensor becomes the one of de Sitter space. This proves the first part of the second sentence of Theorem~\ref{theorem:main}. Notice that it is straightforward to verify that one can reach the same conclusions also for the case of the sphere provided the initial volume satisfies the inequality \eqref{volumeS2}~\footnote{Notice that in this case one could have $K_m(\lambda)\leq K_\Lambda$ everywhere, but all our conclusions still apply in light of the sentence below the formulation of Theorem~\ref{th:boundonmax}.}.  \\

{\bf Physical equivalence with de Sitter space:} We are now ready to show the last part of  Theorem~\ref{theorem:main}, {\it i.e.}~that there are regions of arbitrarily large physical volume which are physically indistinguishable from de Sitter space. By physical indistinguishable we mean that the result of any measurement, with arbitrary but finite accuracy, is the same as done in de Sitter space, at sufficiently large $\lambda$. Let us start the discussion within classical physics and later consider quantum mechanical effects.

\begin{figure}[h] 
\centering
{\includegraphics[scale=.40,angle=90]{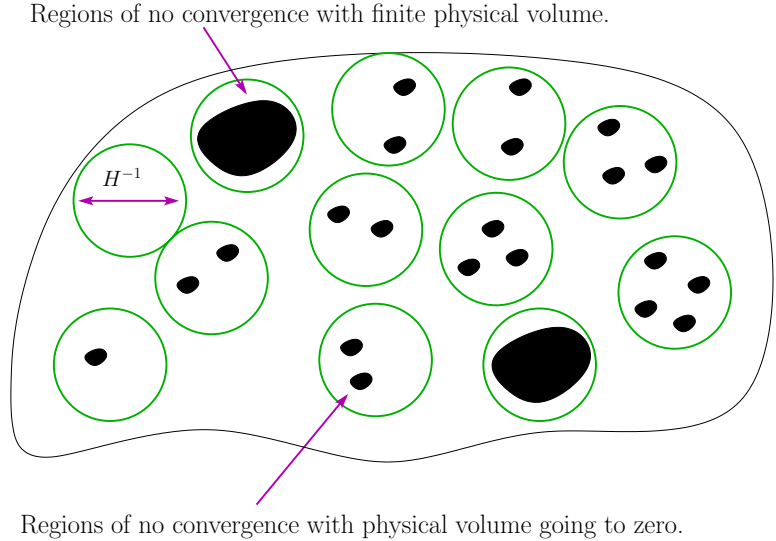}} \quad\quad\quad
\caption{\small Pictorial representation of the surface of MCF at a given $\lambda$ time. The green circles represent regions whose physical size is equal to the Hubble volume. The white regions are extremely close to de Sitter space, while black regions are far from it. Only a fraction of the Hubble patches that goes to zero  can be populated by regions of no-convergence to de Sitter that have finite physical volume, all the other Hubble patches can be populated by regions of non-converge whose physical volume goes to zero.}\label{fig:cow}
\end{figure}

So far, we have shown that for $\lambda \to \infty$, the volume of the region where the stress tensor and the geometric quantities do not converge to de Sitter is at most finite. Given that the physical volume goes to infinity, this means that all quantities converge to de Sitter on average. The most general situation still allowed by our theorem is represented pictorially in Fig.~\ref{fig:cow} at a given $\lambda$. There can be regions of no-convergence to de Sitter with finite physical volume (for example the regions where black holes form\footnote{Static black holes do not really exist in $2+1$ dimensions, but generically crunching singularities will form. We loosely refer to these as black holes.}), and also other regions whose physical volume shrinks to zero that densely populate the whole volume. In particular our results do not imply that there are regions of arbitrarily large volume that {\em pointwise} converge to de Sitter. 

In fact, it is easy to come up with a counterexample to show that pointwise convergence cannot be true in general. Imagine an alien civilisation which, for unknown reasons, wants to prevent pointwise convergence to de Sitter. At a certain time they put one spaceship in each Hubble volume, so that there are no large regions which are close to de Sitter. As time goes on, they divide their spaceships in smaller and smaller spaceships (one can assume all pieces have the same mass density), keeping one piece in each Hubble patch. Notice that they can do this without the need of superluminal travel, so that everything satisfies the energy conditions that we assumed. Since energy is conserved the spaceships become of smaller and smaller mass as time goes on. However in the exact place where each spaceship is sitting there is no convergence to de Sitter, so that one does not have pointwise convergence in any Hubble patch. This shows one does not have in general pointwise convergence without further assumptions.
This counterexample is, however, still quite benign. Given that the spaceships are becoming smaller and smaller, any observation with arbitrary small but finite precision will eventually give the same results as in de Sitter. The solution becomes physically indistinguishable from de Sitter. 

Unfortunately we cannot conclude that this is always what happens. One can imagine that the regions with finite physical volume that do not converge to de Sitter move around in a way that any observer at arbitrarily large $\lambda$, sooner or later, is sensitive to them and realize he/she is not in de Sitter. One cannot exclude this scenario albeit it looks unlikely~\footnote{
It is tempting to say that the regions that converge to de Sitter (up to corrections of vanishing physical volume) will be screened by an event horizon from the other regions that do not converge to de Sitter. However this seems hard to prove. One could imagine that, instead of localised spaceships, the regions of no-convergence with vanishing physical volume are filamentary structures that connect each Hubble patch to far-away regions of finite physical volume that do not converge to de Sitter. Since one does not have pointwise convergence to de Sitter along these structures, one could imagine that these filaments are enough to allow the regions that do not converge to de Sitter to wander around and sooner or later visit any Hubble patch.}.

Fortunately these worrisome scenarios are excluded once quantum mechanics is taken into account. Consider for example the counterexample with the alien spaceships. One cannot define the position of an object with mass $m$ with a precision better than its Compton wavelength $\lambda_c =h/mc$. This corresponds to a maximum density of order $m^4$ in natural units. Since the spaceships go on splitting, the mass of each one goes to zero. This implies, once quantum mechanics is taken into account, that also the energy density goes to zero. When the energy density becomes parametrically smaller than the one of the cosmological constant, the alien spaceships become (pointwise) small perturbations and one has pointwise convergence to de Sitter. The same holds for massless particles: if one has a total energy $E$, all the massless particles have wavelength larger than $1/E$ so that it is impossible to create an energy density higher than $E^4$. (The same logic applies if matter is distributed in filaments or sheets.)  Since the total energy of matter is finite, eq.~\eqref{eq:Tmunubound}, while the volume goes to infinity, the energy in these regions goes to zero~\footnote{Perhaps more simply, one can conclude that since the  energy of each region goes  to zero,  these regions cannot have asymptotically-vanishing physical volume because of quantum mechanics, and so cannot densely populate the whole volume, against the hypothesis.}. Therefore, once quantum mechanics is included, one concludes that the spacetime pointwise converges to de Sitter except for a finite region. Notice that once a sufficiently large region of space is pointwise close to de Sitter, it will become completely insensitive to the rest of space due to the existence of an event horizon. This shows that observers in this region will be in de Sitter {\em forever}.

 \section{Conclusions}
 We proved that a 2+1 dimensional cosmology with a positive cosmological constant, under the assumptions stated at the beginning of Section \ref{sec:theorem}, asymptotically converges to de Sitter, in the sense that there are regions of infinite volume that become closer and closer to de Sitter. This is probably as close as one can get to the notion of a de Sitter no-hair theorem \cite{Wald:1983ky}.
 Starting from a finite initial volume one gets an infinite volume of de Sitter space. This addresses the `initial patch problem': one does not need quasi homogeneous initial conditions on an inflationary Hubble patch for inflation to start. Notice that our arguments are immune from the so-called measure problem: under our assumptions, the probability of having inflation somewhere is one.  Of course the measure problem may return if one tries to estimate how likely inflation is for a set of observers: for instance if we populate the initial surface with a homogeneous density of observers, it may be that most of them will end up in black holes.
 
The natural continuation of this work is to prove the same statements in the 3+1 case. Work is in progress in this direction \cite{inprogress}. One could also try to relax some of the assumptions of the theorem, for instance the one that requires that the initial surface is expanding everywhere, or the approximation of the inflaton potential as a cosmological constant.

 \section*{Acknowledgements}
It is a pleasure to thank Or Hershkovits for collaboration in the earlier stages of this work and for patient mathematical consultancy. 
We also thank Klaus Ecker, Alan Guth, Larry Guth, Matt Kleban, Rafe Mazzeo, Mehrdad Mirbabayi, Richard Schoen and Vicharit Yingcharoenrat for useful conversations. 
LS is partially supported by Simons Foundation Origins of the Universe program (Modern Inflationary Cosmology collaboration) and by NSF award 1720397. AV is partially supported by NSF award DMS-1664683.

\begin{small}
 
\bibliography{references}

\providecommand{\href}[2]{#2}\begingroup\raggedright\begin{thebibliography}{10}

\bibitem{Ijjas:2015hcc}
A.~Ijjas and P.~J. Steinhardt, \emph{{Implications of Planck2015 for
  inflationary, ekpyrotic and anamorphic bouncing cosmologies}},
  \href{http://dx.doi.org/10.1088/0264-9381/33/4/044001}{\emph{Class. Quant.
  Grav.} {\bf 33} (2016) 044001}, [\href{http://arxiv.org/abs/1512.09010}{{\tt
  1512.09010}}].

\bibitem{East:2015ggf}
W.~E. East, M.~Kleban, A.~Linde and L.~Senatore, \emph{{Beginning inflation in
  an inhomogeneous universe}},
  \href{http://dx.doi.org/10.1088/1475-7516/2016/09/010}{\emph{JCAP} {\bf 1609}
  (2016) 010}, [\href{http://arxiv.org/abs/1511.05143}{{\tt 1511.05143}}].

\bibitem{Kleban:2016sqm}
M.~Kleban and L.~Senatore, \emph{{Inhomogeneous Anisotropic Cosmology}},
  \href{http://dx.doi.org/10.1088/1475-7516/2016/10/022}{\emph{JCAP} {\bf 1610}
  (2016) 022}, [\href{http://arxiv.org/abs/1602.03520}{{\tt 1602.03520}}].

\bibitem{barrow1985closed}
J.~D. Barrow and F.~J. Tipler, \emph{Closed universes: their future evolution
  and final state}, {\emph{Monthly Notices of the Royal Astronomical Society}
  {\bf 216} (1985) 395--402}.

\bibitem{Clough:2016ymm}
K.~Clough, E.~A. Lim, B.~S. DiNunno, W.~Fischler, R.~Flauger and S.~Paban,
  \emph{{Robustness of Inflation to Inhomogeneous Initial Conditions}},
  \href{http://dx.doi.org/10.1088/1475-7516/2017/09/025}{\emph{JCAP} {\bf 1709}
  (2017) 025}, [\href{http://arxiv.org/abs/1608.04408}{{\tt 1608.04408}}].

\bibitem{Clough:2017efm}
K.~Clough, R.~Flauger and E.~A. Lim, \emph{{Robustness of Inflation to Large
  Tensor Perturbations}},
  \href{http://dx.doi.org/10.1088/1475-7516/2018/05/065}{\emph{JCAP} {\bf 1805}
  (2018) 065}, [\href{http://arxiv.org/abs/1712.07352}{{\tt 1712.07352}}].

\bibitem{Pretorius:2005gq}
F.~Pretorius, \emph{{Evolution of binary black hole spacetimes}},
  \href{http://dx.doi.org/10.1103/PhysRevLett.95.121101}{\emph{Phys. Rev.
  Lett.} {\bf 95} (2005) 121101},
  [\href{http://arxiv.org/abs/gr-qc/0507014}{{\tt gr-qc/0507014}}].

\bibitem{besse1987einstein}
A.~Besse, \emph{Einstein Manifolds}.
\newblock ~Classics in mathematics. World Publishing Company, 1987.

\bibitem{thurston1997three}
W.~Thurston and S.~Levy, \emph{Three-dimensional Geometry and Topology}.
\newblock No.~v. 1 in Luis A.Caffarelli. Princeton University Press, 1997.

\bibitem{10.2307/2152760}
G.~Perelman, \emph{Manifolds of positive ricci curvature with almost maximal
  volume}, {\emph{Journal of the American Mathematical Society} {\bf 7} (1994)
  299--305}.

\bibitem{gerhardtbook}
C.~Gerhardt, \emph{Curvature Problems}.
\newblock International Press, Boston, 2006.

\bibitem{inprogress}
P.~Creminelli, M.~Kleban, M.~Mirbabayi, L.~Senatore and A.~Vasy, \emph{{in
  progress}}, .

\bibitem{Geroch}
R.~Geroch, \emph{Domain of dependence},
  \href{http://dx.doi.org/http://dx.doi.org/10.1063/1.1665157}{\emph{Journal of
  Mathematical Physics} {\bf 11} (1970) 437--449}.

\bibitem{EH}
K.Ecker and G.~Huisken, \emph{{Parabolic methods for the construction of
  spacelike slices of prescribed mean curvature in cosmological spacetimes}},
  {\emph{Commun.Math.Phys.} {\bf 135} (1991) 595}.

\bibitem{Eardley}
D.~M. Eardley and L.~Smarr, \emph{Time functions in numerical relativity:
  Marginally bound dust collapse},
  \href{http://dx.doi.org/10.1103/PhysRevD.19.2239}{\emph{Phys. Rev. D} {\bf
  19} (Apr, 1979) 2239--2259}.

\bibitem{barrow1986closed}
J.~D. Barrow, G.~J. Galloway and F.~J. Tipler, \emph{The closed-universe
  recollapse conjecture}, {\emph{Monthly Notices of the Royal Astronomical
  Society} {\bf 223} (1986) 835--844}.

\bibitem{E}
K.Ecker, \emph{{On mean curvature flow of spacelike hypersurfaces in
  asymptotically at space-times}}, {\emph{J. Austr. Mat. Soc.} {\bf 55} (1993)
  41}.

\bibitem{Mirbabayi:2018suw}
M.~Mirbabayi, \emph{{Topology of Cosmological Black Holes}},
  \href{http://arxiv.org/abs/1810.01431}{{\tt 1810.01431}}.

\bibitem{bartnik1984}
R.~Bartnik, \emph{Existence of maximal surfaces in asymptotically flat
  spacetimes}, {\emph{Comm. Math. Phys.} {\bf 94} (1984) 155--175}.

\bibitem{Wald84}
R.~M. Wald, \emph{General Relativity}.
\newblock The University of Chicago Press, Chicago, 1984.

\bibitem{Wald:1983ky}
R.~M. Wald, \emph{{Asymptotic behavior of homogeneous cosmological models in
  the presence of a positive cosmological constant}},
  \href{http://dx.doi.org/10.1103/PhysRevD.28.2118}{\emph{Phys. Rev.} {\bf D28}
  (1983) 2118--2120}.

\end{thebibliography}\endgroup


\end{small}

\end{document}